\documentclass{JAC2003}
\usepackage{graphicx}
\usepackage{booktabs}
\usepackage{array}
\usepackage{subfigure}

\begin{document}
\title{STATUS OF HIGHER ORDER MODE BEAM POSITION MONITORS IN 3.9~GHZ SUPERCONDUCTING ACCELERATING CAVITIES AT FLASH\thanks{Work supported in part by the European Commission under the FP7 Research Infrastructures grant agreement No.227579.}\\[-.8\baselineskip]}

\author{P.~Zhang$^{1,2,3,}$\thanks{Now at CERN, Geneva, Switzerland, pei.zhang@cern.ch}, N.~Baboi$^2$, R.M.~Jones$^{1,3,}$\thanks{Now on visit at Yale University, New Haven, CT, U.S.A.}, T.~Flisgen$^{4}$, U.~van~Rienen$^{4}$, I.R.R.~Shinton$^{1,3,}$\thanks{Now at Elekta Limited, Crawley, U.K.}\\
\mbox{$^1$School of Physics and Astronomy, The University of Manchester, Manchester, U.K.}\\
\mbox{$^2$Deutsches Elektronen-Synchrotron (DESY), Hamburg, Germany}\\
\mbox{$^3$The Cockcroft Institute of Accelerator Science and Technology, Daresbury, U.K.}\\
\mbox{$^4$Institut f\"ur Allgemeine Elektrotechnik, Universit\"at Rostock, Rostock, Germany}}

\maketitle


\begin{abstract}
Higher order mode (HOM) beam position monitors (BPM) are being developed for the 3.9~GHz third harmonic superconducting accelerating cavities at FLASH. The transverse beam position in a cavity can be determined utilizing beam-excited HOMs based on dipole components. The existing couplers used for HOM suppression provide necessary signals. The diagnostics principle is similar to a cavity BPM, but requires no additional vacuum instruments on the linac. The challenges of HOM-BPM for 3.9~GHz cavities lie in the dense HOM spectrum arising from the coupling of the majority HOMs amongst the four cavities in the cryo-module ACC39. HOMs with particularly promising diagnostics features were evaluated using a spectrum analyzer and custom-built test electronics with various data analysis techniques, data reduction was focused on. After careful theoretical and experimental assessment of the HOM spectrum, multi-cavity modes in the region of 5~GHz were chosen to provide a global position over the complete module with superior resolution ($\sim$20~$\mu m$) while trapped modes in the 9~GHz region provide local position in each cavity with comparable resolution ($\sim$50~$\mu m$). A similar HOM-BPM system has been planned for the European XFEL 3.9~GHz module which encompasses eight cavities. This paper reviews both the current status and the future prospects of HOM-BPMs in 3.9~GHz cavities at FLASH.
\end{abstract}


\section{Introduction}
Charged particle beams excite wakef{}ields when passing through accelerating cavities \cite{wake-1}. These wakef{}ields may be decomposed into higher order modes (HOM). The long-range wakef{}ield is focused on. Particles trailing those ahead will be af{}fected by these electromagnetic f{}ields, and hence the beam quality may be deteriorated. Therefore, for the acceleration of high intensity beams, suppressing wakef{}ields is an important consideration for many accelerators \cite{wake-1}. The longitudinal wakef{}ields are unavoidable, but their ef{}fects can be partly compensated by adjusting the RF input power along the train of bunches accordingly. Transverse wakef{}ields on the other hand, can only be excited by of{}f-axis beams in cylindrically symmetric structures, thus can be minimized by positioning the beam within the electric center of the cavity. Moreover, transverse wakef{}ields carry beam position information, and this enables position diagnostics inside existing cavities without the need for additional vacuum instrumentation, thus a higher order mode beam position monitor (HOM-BPM).

\begin{figure}[h]\centering
\includegraphics[width=0.49\textwidth]{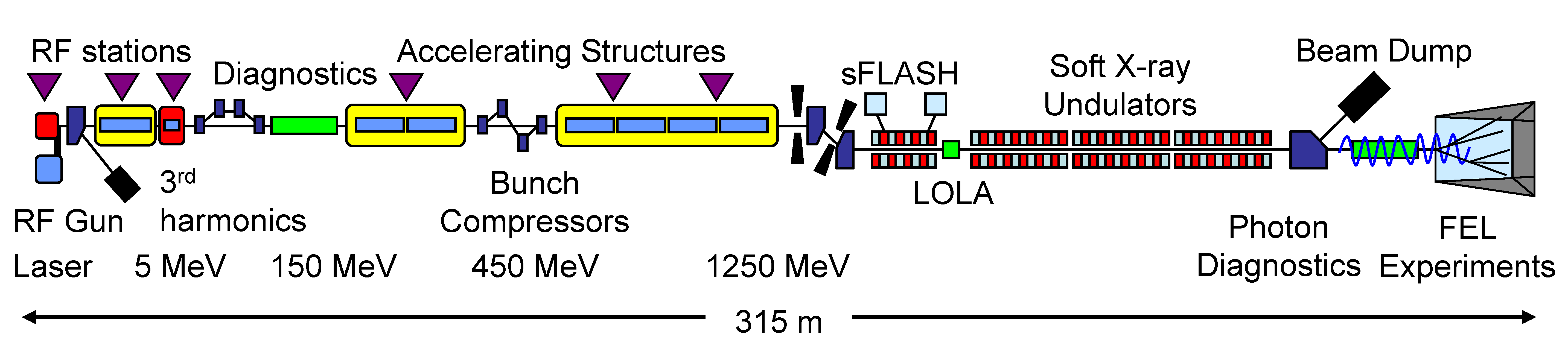}
\caption{Schematic layout of the FLASH facility \cite{flash-1}.}
\label{flash-layout}
\end{figure}
FLASH \cite{flash-1} is a free-electron laser (FEL) facility at DESY (Hamburg, Germany) providing ultra-short radiation pulses with an unprecedented brilliance (Fig.~\ref{flash-layout}). It is a user facility for fundamental light source applications and test facility for European XFEL \cite{xfel} and ILC \cite{ilc}. A photoelectric gun generates trains of electron bunches. The bunch train is up to 800~$\mu s$ long and has a repetition rate of 10~Hz. The spacing between bunches in the train is up to 1~MHz and the charge of each bunch is between 100~pC and 1~nC. The electrons are accelerated to maximum energy of 1.2~GeV by 7 accelerating modules, each containing 8 superconducting (sc) 1.3~GHz TESLA-type cavities. Four third harmonic sc cavities working at 3.9~GHz are used to linearize the longitudinal phase space of the bunch \cite{acc39-0}. In the undulator section the beam produces a laser-like coherent FEL radiation with a wavelength between 4 and 45~nm. 

HOM-BPMs have been previously built for 1.3~GHz cavities at FLASH \cite{tesla-hombpm-1} and it is planned to build similar system for 3.9~GHz ones. However, signif{}icant challenges arise due to the prominent features of the 3.9~GHz cavities. This paper reviews the current status and the future prospects of HOM-BPMs in 3.9~GHz cavities at FLASH \cite{thesis}.

In a periodic structure with cylindrical symmetry, transverse wakef{}ields $\mathbf{W_{\perp}}$ is dominated by dipole modes as (for positions behind the bunches: $s>0$) 
\begin{equation}
\mathbf{W_\perp} \simeq \left(\hat{r}cos\theta - \hat{\theta}sin\theta\right) r' c \sum_n^\infty \left(\frac{R}{Q}\right)_{n} sin\frac{\omega_{n}s}{c},
\label{eq:prox-wake-2}
\end{equation}
where $\hat{r}$ and $\hat{\theta}$ are the unit vectors in $r$ and $\theta$ directions. The transverse position of the excitation particle is denoted as $(r',\theta')$ where $\theta'$ has been set to $0$ because of the cylindrical symmetry. 
$\left(R/Q\right)_{n}$ and $\omega_{n}$ are the ratio of shunt impedance to quality factor and radian frequency of the $n^{th}$ dipole eigenmode. The $(\hat{r}cos\theta - \hat{\theta}sin\theta)$ term describes the polarization of the dipole mode in the transverse plane. There are two polarizations of each dipole mode perpendicular to each other. The dipole transverse wakef{}ield is of particular interest, since it has a linear dependence on of{}fset $r'$ of the particle inducing the wakef{}ield, regardless of the observation position. This makes it possible to determine the transverse beam position inside the cavity by examining the beam-excited dipole signal.

\begin{figure}[h]\centering
\includegraphics[width=0.48\textwidth]{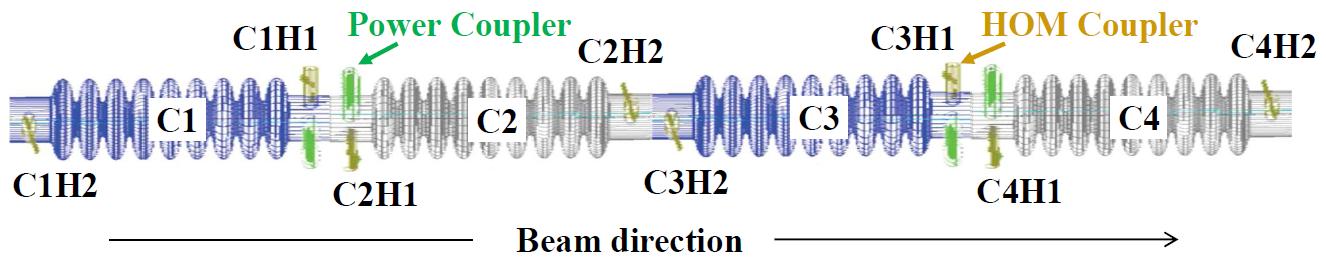}
\caption{Schematic of 4 cavities within ACC39 module.}
\label{cavity-cartoon}
\end{figure}

The 3.9~GHz module, ACC39, is composed of four third harmonic sc cavities illustrated in Fig.~\ref{cavity-cartoon}. Each cavity is equipped with two HOM couplers to suppress wakef{}ield.

By design \cite{acc39-0}, the 3.9~GHz cavity has two features in terms of wakef{}ields compared to its 1.3~GHz counterpart. F{}irstly, the wakef{}ields in the 3.9~GHz cavity are stronger as the iris radius is signif{}icantly smaller: 15~mm in the 3.9~GHz cavity compared to 35~mm in the 1.3~GHz one. From scaling considerations, transverse wakef{}ields $W_\perp$ grow as $W_\perp \sim \mathrm{(frequency)}^{-3}$. Secondly, HOM spectra of 3.9~GHz cavities are signif{}icantly more complex. The main reason for this is that, unlike the TESLA cavity case, the majority modes are above the cutof{}f frequencies of the beam pipes. This allows most of the modes from each independent cavity to couple through to adjacent cavities. In this case, most modes reach all eight HOM couplers, which facilitates the distributed HOM damping \cite{acc39-0}. This leads to almost of the modes, apart from select bands, in all four cavities in the module being coupled and hence they are particularly sensitive to geometrical tolerances.  This gives rise to the complexity in HOM spectra, which makes the realization of HOM-BPM for 3.9 GHz cavities challenging.


\section{HOM-BPM Development at FLASH}

\subsection{Feasibility Studies by a Spectrum Analyzer}
\begin{figure}[h]\centering
\includegraphics[width=0.48\textwidth]{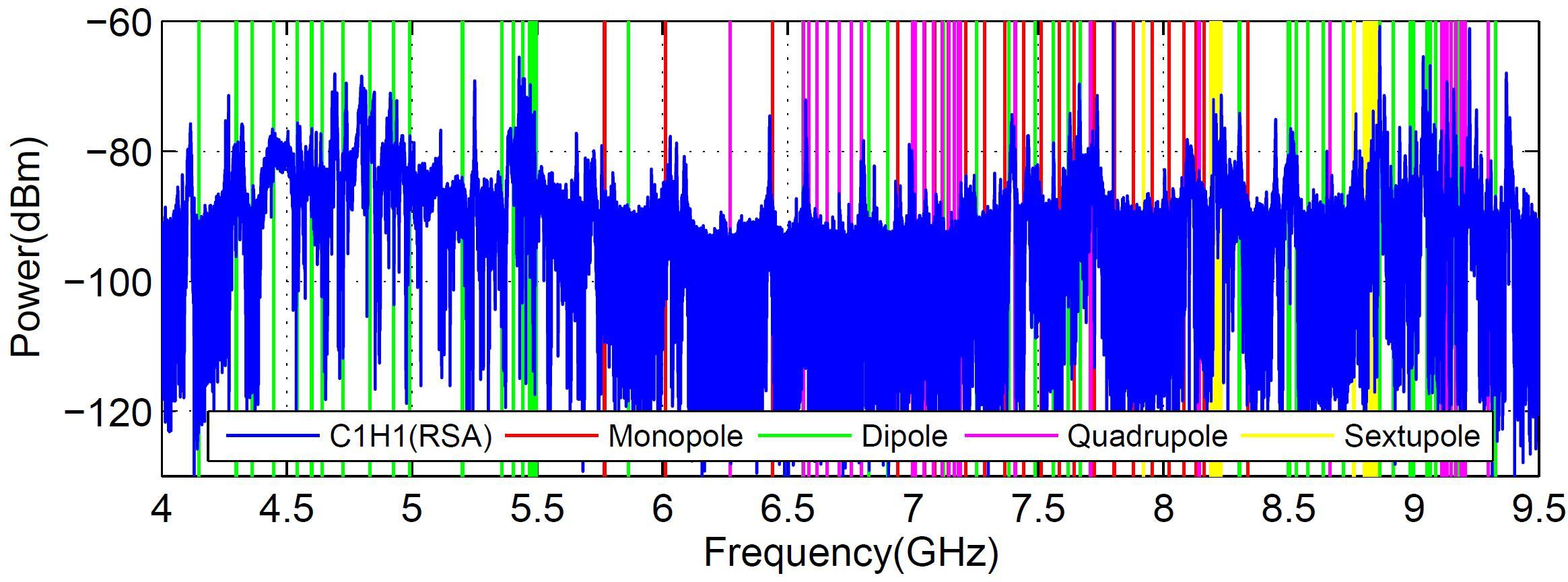}
\caption{Beam-excited spectrum measured from coupler C1H1 by a spectrum analyzer. The vertical lines indicate simulation results \cite{thesis}.}
\label{rsa-hom}
\end{figure}

\begin{figure}[h]\centering
\includegraphics[width=0.48\textwidth]{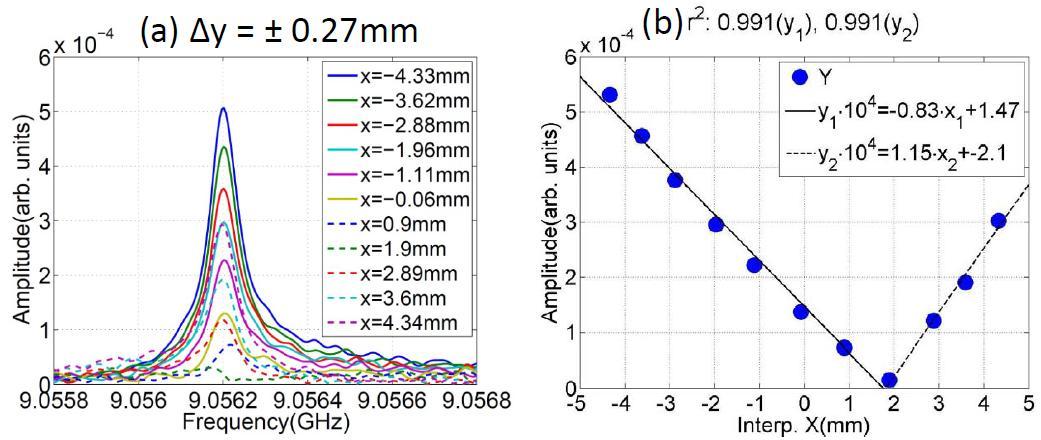}
\caption{Dipole dependence on transverse beam of{}fset.}
\label{dep}
\end{figure}

Prior to developing electronics for HOM-BPM, it is essential to characterize the HOMs and understand their behavior relating to beam of{}fset. To this end, extensive simulations and spectra measurements have been made. A beam-excited spectrum measured from one HOM coupler using a spectrum analyzer is shown in Fig.~\ref{rsa-hom} along with simulation results. A typical dipole-like behavior of one example HOM is shown in Fig.~\ref{dep}.  

After careful theoretical and experimental assessment of the HOM spectrum, we ascertained that there are three signif{}icant regions of interest useful for beam position diagnostics \cite{baboi-1,acc39-hombpm-1}. The f{}irst region at 4.5--5.5~GHz contains modes which are coupled from one cavity to the next within the ACC39 module. The next region at $\sim$9~GHz is the f{}ifth dipole band containing trapped cavity modes. The third region at $\sim$4~GHz is the localized dipole beam-pipe modes. Each region has its own merits. The coupled cavity region allows superior position accuracy, whereas the trapped cavity modes or localized beam-pipe modes allow diagnostics on a localized cavity or beam pipe, but with a reduced accuracy compared to the coupled one.

The decision of modes suitable for HOM-BPM cannot be made without a position resolution study for each option. Therefore, a custom-built test electronics was developed by FNAL and DESY to overcome the resolution limitation of the spectrum analyzer. 

\subsection{Resolution Studies by Test Electronics}
\begin{figure}[h]\centering
\includegraphics[width=0.45\textwidth]{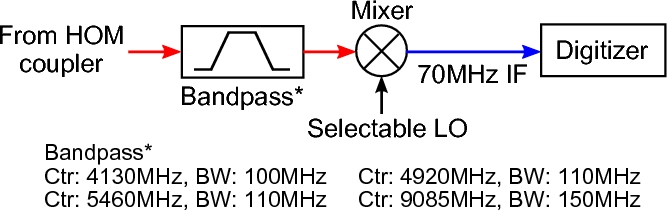}
\caption{Schematics of the test electronics.}
\label{box}
\end{figure}

\begin{figure}[h]\centering
\includegraphics[width=0.47\textwidth]{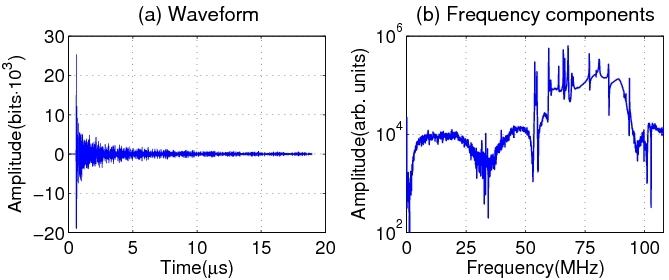}
\caption{An example waveform and its Fourier transform.}
\label{wfm-fft}
\end{figure}

The test electronics was designed to have the f{}lexibility to study the modal options of interest as well as accommodating the large frequency bandwidth (BW) of these options. Its simplif{}ied block diagram is shown in Fig.~\ref{box}. One of four bandpass f{}ilters can be connected into the chain to study the various modal options. After f{}iltering the signal is mixed with a selectable local oscillator (LO) to an intermediate frequency (IF) of ca.~70~MHz, and subsequently digitized. Both the LO and digitizer clock are locked to FLASH master oscillator for correct phasing. An example digitized signal with RF center frequency of 9066~MHz is shown in Fig.~\ref{wfm-fft} along with its frequency components. 

All regions of interest in the HOM spectrum have been carefully assessed by using the test electronics \cite{acc39-hombpm-2}. The resolutions in predicting transverse beam positions are listed in Table~1 by using these modal options. Based on these results, the specif{}ication of the HOM-BPM for the 3.9~GHz cavities were f{}inalized \cite{baboi-2}. Dedicated electronics will be installed on two HOM couplers using coupled modes centered at 5440~MHz with 100~MHz bandwidth to provide high-resolution position over the entire module. In addition, six HOM couplers will be equipped with electronics centered at 9060~MHz with 100~MHz bandwidth to facilitate local position measurements inside each cavity. Since not all modes in the f{}ifth dipole band are localized \cite{acc39-hombpm-1}, the frequency band has been carefully chosen in order to mitigate contaminations of coupled modes. 
\begin{table}[h]
\centering
\caption{Position resolutions for various modal options.}
\begin{tabular}{cccc}
\toprule
\textbf{Mode} & \textbf{f}(GHz) & \textbf{Resolution} & \textbf{Position type}\\ 
\midrule
Beam-pipe & $\sim$4.1 & $\sim$100~$\mu m$ & beam pipe\\
\midrule
1$^{st}$ dipole band & $\sim$4.9 & $\sim$30~$\mu m$ & module \\
\midrule
2$^{nd}$ dipole band & $\sim$5.4 & $\sim$20~$\mu m$ & module \\
\midrule
5$^{th}$ dipole band & $\sim$9.05 & $\sim$50~$\mu m$ & single cavity \\
\bottomrule
\end{tabular}
\label{res-table}
\vspace*{-3mm}
\end{table}

\subsection{Current Status}
HOM-BPM electronics are currently being built by FNAL in collaboration with DESY. The commissioning and the integration to FLASH control system will take place in late 2013 when FLASH restarts its operation \cite{flash-1}. 

As also observed in the previous TESLA 1.3~GHz cavity case, the calibration of the HOM-BPM was only stable for 1--2 days. This may be due to phase drifts or even drifts of the HOM frequencies. This issue need to be understood and is planned to be studied for both 1.3~GHz and 3.9~GHz HOM-BPM system. This is essential for a stable operation of an online beam diagnostics instrument. Based on the results obtained so far, it has been planned to build similar HOM-BPM system for the European XFEL third harmonic module consisting of eight 3.9~GHz cavities. 


\section{Final Remarks}
Theoretically, in cylindrically symmetric structures, transverse wakef{}ields are only excited by of{}f-axis beams, but in practice, they are always excited and present. This is due to the cavity being non-symmetric attributed to couplers installation and cell-misalignment from fabrication tolerances, and also added by the fact that cavities are misaligned inside the cryo-module. In this case, one can minimize the detrimental ef{}fect from transverse wakef{}ields by aligning the beam to the cavity axis. From here the idea came to improve the beam quality by monitoring the transverse wakef{}ield. As we know, the dominant components of the transverse wake are dipole modes, whose amplitude depends linearly on the bunch of{}fset. By monitoring dipole components of the wake, one can not only reduce the f{}ield excitation, but measure the beam of{}fset as well. This means a direct access to the beam properties inside the cryo-module, where only few diagnostics components are normally present. In addition, this requires no vacuum components to be installed on the linac, as the existing HOM couplers provide necessary signals. A limited suite of monitoring electronics and associated software is all that is necessary to implement a diagnostic system -- which is of course comparatively inexpensive.

Beam position information is concealed in the beam-excited inter-cavity coupled modes, which constitute a complex HOM signal radiated to the pickups. Extraction of this information requires advanced data analysis methods. Linear regression along with dimension reduction techniques has been successfully applied \cite{acc39-hombpm-1,acc39-hombpm-2}. Improved position resolutions and computation ef{}f{}iciencies have been achieved with a noise reduction as a byproduct \cite{acc39-hombpm-3}.

For multi-bunch operations, a non-degenerate position resolution can be achieved when diagnosing the f{}irst bunch in the train by using a proper time window on the HOM signals. Due to an overlap of HOM signals with that excited from the following bunches, degenerate position resolution can be foreseen for the trailing bunches in the train \cite{acc39-hombpm-2}.



\end{document}